\documentclass[pre,floatfix,aps,twocolumn]{revtex4-1}

\usepackage{amssymb}
\usepackage{subfigure}
\usepackage[utf8]{inputenc}
\usepackage{graphicx}
\usepackage{amsmath}
\usepackage{amsfonts}
\usepackage{bm}
\usepackage{color}
\usepackage{hyperref}

\hypersetup{colorlinks=true, citecolor=blue,
  urlcolor=black,
  pdfcreator={pdflatex},
}

\begin{document}
\title{A minimal statistical-mechanical model for multihyperuniform
patterns in avian retina}

\author{Enrique Lomba$^{1}$, Jean-Jacques Weis$^3$, Leandro Guisández$^{1,2}$,
  and Salvatore Torquato$^{4,5}$}
\affiliation{$^1$Instituto de Qu\'{i}mica F\'{i}sica Rocasolano,
CSIC, Calle Serrano 119, E-28006 Madrid,
Spain\\
$^2$IFLYSIB (UNLP, CONICET), 59 No. 789, B1900BTE La Plata, Argentina\\
$^3$Université de Paris-Saclay, Laboratoire de Physique Théorique, Bâtiment 210, 91405 Orsay Cedex, France\\
$^4$Department of Chemistry, Princeton University, Princeton, New Jersey 08544, USA\\
$^5$Princeton Institute for the Science and Technology of Materials, Princeton University, Princeton, New Jersey
08544, USA
}

\begin{abstract}
Birds are known for their extremely acute sense of vision. The very
peculiar structural distribution of five different types of
cones in the retina underlies this exquisite ability to sample light.
It was recently  
found that each  cone population as well as their total population display a disordered pattern
in which long wave-length density fluctuations vanish [Jiao et al., Phys. Rev. E,
{\bf 84}, 022721 (2014)]. This property, known as hyperuniformity  is
also present in   perfect crystals.  In situations
  like the avian retina in
  which both the global structure and that of each component
  display hyperuniformity, the system is said to be
  multi-hyperuniform. In this
work, we aim at devising a minimal statistical-mechanical model that
can reproduce the main features of the spatial distribution of photoreceptors in avian
retina, namely the presence of disorder, multi-hyperuniformity and local
hetero-coordination. This last feature is key to avoid local
clustering of the same type of photoreceptors, an undesirable feature
for the efficient sampling of light. For this purpose we formulate  a 
simple model that definitively exhibits  the required structural
properties, namely an equimolar three-component mixture
(one component to sample each primary color, red,
  green, and blue) of
non-additive hard disks to which a long-range logarithmic repulsion is
added between like particles.  A Voronoi
analysis of our idealized system of photoreceptors shows that the space-filling Voronoi polygons interestingly display a rather uniform area
distribution, symmetrically centered around that of a regular
lattice, a structural property also found in human
retina. Disordered multi-hyperuniformity offers an alternative to
generate photoreceptor patterns with minimal long-range
concentration and density fluctuations. This is the key to overcome the
difficulties in devising an  efficient visual system in which
crystal-like order is absent. 
\end{abstract}

\maketitle
\section{Introduction}
Sampling light is one of the essential activities that enables the
interaction of living organisms with the surrounding environment. From
simple devices such as the stigma that provides ``vision'' in certain
classes of microalgae \cite{Hegemann1997}, to the sophisticated
compound eyes of arthropods \cite{Ready1976,Lubensky2011} living organisms have
developed increasingly efficient ways to map visual information from the
external world onto signals that can be processed by their cognitive
systems. The case of arthropod eyes is
particularly interesting. It is known from classical sampling theory
\cite{Petersen1962} that an optimal sampling of light can be achieved
by an hexagonal array of photodetectors. This is actually the pattern
adopted by ommatidia (the optical units forming a compound eye) in
arthropods. Compound eyes are imaging systems with
low aberration, wide-angle field of view and infinite field
depth\cite{Song2013}. These properties have  motivated intense
research into the development of bionic compound eyes intended for
small robots\cite{Zheng2019} or sensors for digital
cameras\cite{Song2013}. 

When it comes to vertebrates, with the
exception of some teleost fish \cite{Raymond2004,jiao2014} and some
reptiles \cite{Dunn1966,jiao2014}, the situation is different and
structural disorder in photoreceptor patterns is the general trend. In
this connection, birds are in a class of their own. They possess  one of
the most elaborate visual systems among vertebrates. In avian retina,
one can find five
different types of cones, \cite{Ruggeri2010,Toomey2017}, one type
for luminance detection and the remaining four building a tetrachromatic color
sensing device covering wavelengths from red to ultraviolet
\cite{WITHGOTT2000}. In contrast with the regular shape of ommatidia
in insects, photoreceptors in bird retina are polydisperse, in size
and number \cite{Goldsmith1984,Moore2012}. This variation provides an
adaptative advantage: changing the relative numbers and even
pigmentation of the cones bird species can have visual capabilities
adapted to different habitats (sea birds have high density of red/yellow
cones for hazy conditions, nocturnal birds have a extremely high
density of luminance cones, ...). However, 
polydispersity is known to
frustrate crystallization \cite{Frenkel2006}, so an alternative
to the regular hexagonal pattern of arthropod
eyes is needed if we want to preserve a good sampling of light. In
this connection, Jiao and coworkers \cite{jiao2014} found that  the spatial distribution of photoreceptors in chicken retina
retained some ``hidden order'' reminiscent of crystalline
patterns. Namely, they found that long-range density and concentration
fluctuations were vanishingly small. This feature can be quantified
by means of two intimately connected structural properties. In two
dimensions, we have first the number variance of cones over a sampling area of radius R,
defined as $\sigma^2_N(R) = \langle N^2\rangle_R -  \langle
N\rangle_R^2$, where  $N$ is the number of cones contained in the
sample area and $\langle\ldots\rangle_R$ denotes the
average over a certain number of sampling areas. In
Ref.~\cite{jiao2014}, it was found that this quantity obeys the following large-$R$ asymptotic scaling
\begin{equation}
\sigma^2_N(R) \propto R
\label{sigma2}
\end{equation}
in the plane. This  is one of the possible scalings of hyperuniform systems, also  characteristic of
crystalline-like order in two dimensions (class I following Ref.~\cite{Torquato2016}). Secondly, it is known that density
fluctuations in Fourier space are directly related to the  structure
factor. This is defined for
a set of points/particles with number density $\rho$ by
\begin{equation}
S({\bf Q}) = 1+ \rho \tilde{h}({\bf Q}),
\label{sq}
\end{equation}
where ${\bf Q}$ is the wave vector, $\tilde{h}({\bf Q})$ is the
spatial Fourier transform of $h(r) = g_2(r) -1$, being $ g_2(r)$ the pair
distribution function of the point/particle configuration. It is
possible to show\cite{Torquato2016} that for a system satisfying
Eq.~(\ref{sigma2}) in two dimensions then
\begin{equation}
S(Q) \propto Q^\alpha\;\; (Q\rightarrow 0)
\label{limQ}
\end{equation}
with $\alpha > 0$. Since
Eq.~(\ref{sigma2}) holds for each cone distribution, then we will have
a relation like (\ref{limQ}) for the structure factor computed from
each cone pattern, as was found by Jiao and coworkers~\cite{jiao2014},
i.e.
\begin{equation}
\lim_{Q\rightarrow 0} S_{ii}(Q) = 0
\label{multi}
\end{equation}
for each cone type $i$.  This implies that 
  density fluctuations of the corresponding point patterns will 
  vanish for long wavelengths, i.e. when ${Q\rightarrow 0}$. The same
  applies to the overall point pattern. This property was termed in
  Ref.~\cite{jiao2014} as ``multi-hyperuniformity''.

Interestingly, since Torquato and Stillinger \cite{Torquato2003}
introduced the concept of hyperuniformity and stressed its
significance in structurally disordered materials, such exotic 
 ``states of matter'' have been found in a wide variety of
systems. A partial list of examples include
amorphous dielectric networks
with large and complete photonic
band gaps \cite{Florescu2009,Froufe-Perez2017}, dense transparent
disordered media \cite{Leseur2016},
   the enhanced
pinning of vortices in  arrays in superconductors\cite{Thien2017},
certain composites with desirable transport, dielectric and fracture 
properties  \cite{Zh16b,Ch18,Xu17,Wu17}, sand piles and other
avalanche models \cite{Dickman2015,Garcia-Millan2018}, driven nonequilibrium
granular and colloidal systems \cite{He15,We15,Tj15} and even
immune system receptors  \cite{Ma15} all have in common the presence
of hyperuniformity. 

One might ask why hyperuniformity plays such a crucial role
in the quality of vision in birds ?. 
 As mentioned, the optimal
sampling configuration of photoreceptors corresponds to a fully regular hexagonal
arrangement. Hyperuniformity prevents long-wavelength fluctuations in
the photoreceptor density  (or concentration of different species)
that would be otherwise be 
present in a structurally disordered configuration of
photoreceptors.   The presence of such
fluctuations is certainly not a desirable property for an
accurate image representation. Fully regular
arrangements such as the hexagonal patterns of ommatidia are
hyperuniform, but in the case of bird retina, crystal-like order 
 is preempted by polydispersity. Thus 
hyperuniform patterns might well be a good compromise
solution. Multi-hyperuniformity will guarantee the same sampling
quality for each type of photoreceptors and aids in ensuring local
hetero-coordination, 
which is key to prevent the unwanted 
clustering of same color photoreceptors. 

After these considerations, it is our aim to build a minimal
statistical mechanical model that can reproduce the main characteristics
of the photoreceptor distribution. These are, in addition to disorder,
on one hand
multi-hyperuniformity, and on the other local hetero-coordination. By
this we mean that photoreceptors of the same type should not be
allowed to cluster together if color sensitivity is to be uniformly
distributed on space. In fact, in Ref.~\cite{Lomba2018} it was shown
that a system can be multi-hyperuniform and display a strong degree of
clustering (chain formation). From pictures of actual chicken
cone distributions (see Figure 1 in Ref.~\cite{Kram2010}) it is
readily apparent that cones of different types tend to cluster
together, i.e. their spatial distribution displays
hetero-coordination. 

The findings  of Ref.~\cite{Lomba2018} suggest
that a mixture with logarithmic long-range repulsions and non-additive
hard-core volume exclusions can display the sought
characteristics. Strictly speaking the model in question was a
two-dimensional Coulomb plasma. Interestingly, in Ref.~\cite{jiao2014}
it was found that the structure factor derived from photoreceptor
patterns displays a small wave number decay consistent with $\sim
Q$ (or  $\sim Q^2$ when fitted into a multiscale packing
model~\cite{jiao2014}).  Strictly two-dimensional
Coulomb plasmas are known to have structure factors that decay
quadratically with the wavenumber as $Q\rightarrow 0$ \cite{Caillol1982}.
 Obviously, here the
logarithmic repulsion is to be thought of as an effective interaction
between   photoreceptors. In order to properly account for the
presence of hetero-coordination, both the long-range and the
short-range hard core repulsions have to be non-additive. 

Additionally, 
we will see that a Voronoi
analysis of the disordered hyperuniform patterns further illustrates
the hidden connection between these and the fully ordered crystal
structures. The area distribution of Voronoi polygons
  is relatively uniform and centered around that of a crystal like
  pattern. This uniformity, also found in the Voronoi tessellation of
photoreceptors in human retina \cite{Legras2018}, is in our case the
result of the presence of a long-ranged monotonic repulsive interaction.

\section{Model and methods}
As mentioned, our 
minimal model of ``retina'' consists of three classes of photoreceptors,
(red-green-blue=RGB) in which, following Ref.~\cite{Lomba2018}, interactions will be defined in terms of a
purely repulsive logarithmic potential.  
In addition, in order to guarantee hetero-coordination
from moderate to high densities, the particles will have a 
hard-core volume exclusion defined by a hard-disk diameter $\sigma$,
with unlike particles having a distance of minimum approach
$(1+\Delta)\sigma$, with $\Delta < 0$. From Ref.~\cite{Lomba2018}, we
know that $\Delta >0$ induces the formation of stable clusters of like particles due to the
combination of long-range like particle repulsions and an effective short range
attraction between like particles due to volume effects.
It is worth stressing that  our 
``minimal model'' in which for computational simplicity the number of
components is reduced to the minimum, three. One can straightforwardly extend
the model to four (cyan-magenta-yellow-black=CMYK) or five types
(including the luminance cones as in bird retina) of photoreceptors. No
significant qualitative difference in the results is to be expected
from the consideration of a larger number of photoreceptor types. 

The
net interaction between  particles of type $i$ and $j$ can be
explicitly written as
\begin{equation}
  \beta u_{ij}(r) = \left\{
  \begin{array}{cc}
    \infty & \mbox{if}\; r < (1+\Delta(1-\delta_{ij}))\sigma \\
    - \gamma_{ij} \log r/\sigma
    \label{uijlog} & \mbox{if}\; r \ge (1+\Delta(1-\delta_{ij}))\sigma
  \end{array}
  \right.
\end{equation}
 where  $\gamma_{ij}$ is an effective coupling parameter, and
 $\delta_{ij}$ is Kronecker's symbol. Our minimal model is fully 
 symmetric, with $u_{ii}=u_{11}$ $\forall i$, and $u_{ij}=u_{12}$
 $\forall i\neq j$. For the logarithmic repulsion the coupling
 parameter is expressed as
\begin{equation}
\gamma_{ij}=(\lambda+(1-\lambda)\delta_{ij})\Gamma
\label{Gama}
\end{equation}
 with $0\le \lambda\le 1$. The parameter $\lambda$ controls the
 non-additivity of the long-range interactions, and we will see it
 determines whether the system displays multi-hyperuniformity or not. 

Now, from our study on binary mixtures in
Refs.\cite{Lomba2017,Lomba2018} we know that disordered systems with long-ranged
repulsive interactions, whose small wavenumber scaling in Fourier
space follows 
\begin{equation}
\lim_{Q\rightarrow}\beta\tilde{u}_{ij}(Q)\propto Q^{-\alpha}
\label{limuQ}
\end{equation}
with $\alpha >0$ will exhibit hyperuniformity. In Ref.~\cite{Lomba2018} we found the conditions
that cross interactions must fulfill for a binary system to be
multi-hyperuniform. Here we extend our analysis, based on the
Ornstein-Zernike (OZ) theory
for mixtures, to multi-component systems. A detailed presentation can
be found in the Supplementary Information. Our key result, here is
that a n-component system in which the   small wavenumber behavior
of the particle-particle interactions follows (\ref{limuQ}), will be
multi-hyperuniform --i.e. comply with Eq.~(\ref{multi})-- if
\begin{equation}
\lim_{Q\rightarrow 0}|{\bf \tilde{u}}(Q)|  \neq  0
\label{multih}
\end{equation}
where $|\ldots |$ denotes a matrix determinant, and the elements of the
matrix ${\bf \tilde{u}}(Q)$ are the Fourier transform of the
species-species interactions, $u_{ij}(r)$. It can be shown that a
sufficient condition for Eq.~(\ref{multih}) to be fulfilled is that 
\begin{equation}
\lim_{Q\rightarrow
  0}\left[\tilde{u}_{ii}(Q)\tilde{u}_{jj}(Q)-\tilde{u}_{ij}(Q)^2
\right]\neq 0,
\label{bt}
\end{equation}
which actually means that cross interactions must {\bf not} comply
with the Lorentz-Berthelot mixing rules in the long wavelength limit. This we had already found for
binary mixtures in Ref.~\cite{Lomba2018}. In practice, for our model
system this means that $\lambda < 1$. Here we will simply set
$\lambda=0$ which reduces cross interactions to bare hard-core
repulsions.

   \begin{figure*}[t]
   \center
   \subfigure[$\rho\sigma^2=0.2$]{\includegraphics[width=7cm,clip]{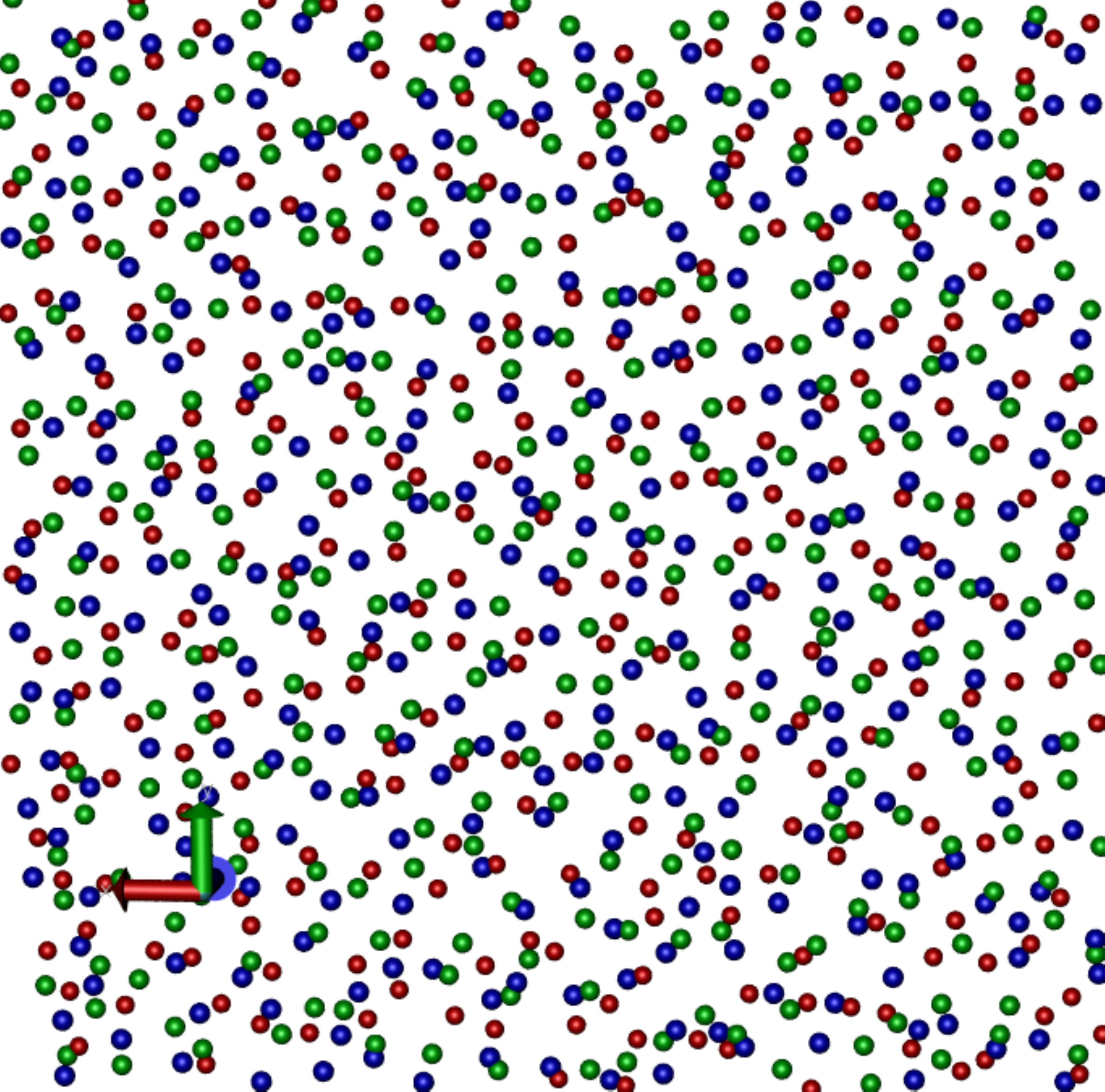}\label{sn3c02}}\hspace{1cm}
   \subfigure[$\rho\sigma^2=0.8$]{\includegraphics[width=7cm,clip]{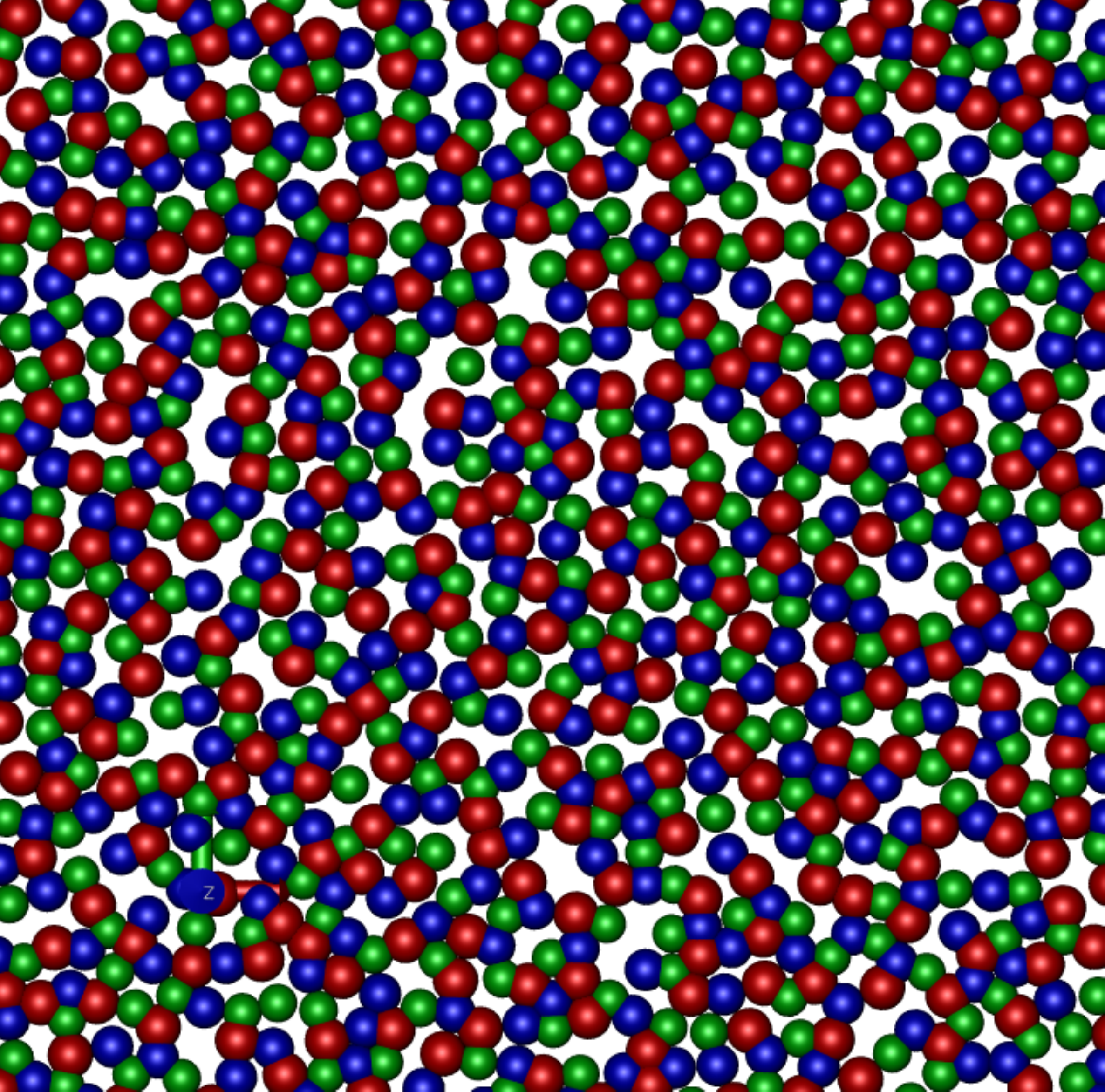}\label{sn3c08}}

   \caption{Snapshots of Monte Carlo configurations of our three
     component minimal model of retina with $rgb$ receptors (as shown
     in the Figure). The interaction is defined by a coupling $\gamma
     = 5$ and a non-additivity parameter $\Delta=-2$.\label{snap3c}}
 \end{figure*}

The low-$Q$ asymptotics of the structure factor when all densities are
identical ($\rho_i=\rho/3$ $\forall i$) simplifies considerably. A detailed derivation can
be found in the supplementary information based on the low-$Q$
expansion of the OZ equation.   In our particular case, given the symmetry of
the interactions and compositions and setting $\lambda=0$, from
Eq.~(S.11) in the SI the limiting
 behavior of the partial structure factors reduces to
\begin{eqnarray}
 \lim_{Q\rightarrow 0}S_{ii}(Q)&=&  Q^2/(2\pi\rho\Gamma), \;\forall i\nonumber\\
 \lim_{Q\rightarrow 0}S_{ij}(Q)&=& \rho
                                   \tilde{c}_{ij}^R(0)Q^4/(2\pi\rho\Gamma)^2,
                                   \;\forall i\neq j.
\label{sijq0}
\end{eqnarray}
with $\tilde{c}_{ij}^R(Q)$ being the Fourier transform of the short
range component of the direct correlation function (cf SI for further
details), which is non-zero and finite as $Q\rightarrow 0$.  When considering mixtures, it
is important to monitor the  global hyperuniformity using the 
the number-number structure factor. This is simply the net structure
factor given by Eq.~(\ref{sq}) where the pair distribution function is
computed using all particle types. In
practice it can be also computed from the addition of the partial
structure factors as
\begin{equation}
  S_{NN}(Q)=\sum_{i,j} S_{ij}(Q).
    \label{snn}
\end{equation}
From Eq.~(\ref{sijq0}) we then have
\begin{equation}
\lim_{Q\rightarrow 0}S_{NN}(Q) =  3Q^2/(2\pi\rho\Gamma)+bQ^4. 
\label{snnq0}
\end{equation}
where $b=3\rho\tilde{c}_{ij}^R(0)/(2\pi\rho\Gamma)^2$.

 The systems studied in this work have been analyzed  using an
integral equation approach based on the OZ equation,
Eq.~(S.1), with a Reference Hypernetted Chain (RHNC) closure
(Eq.~(11) in Ref.~\cite{Lomba2017}). We refer the
reader to \cite{Lomba2017} for further details on the numerical
approach to solve this equation. We have also performed extensive canonical (NvT) Monte
Carlo simulations, in which the energy of the periodic system is
evaluated using the Ewald technique with conducting boundary
conditions\cite{Leeuw1982,Lomba2017}. Computational details of the
simulations are identical to those of Ref.~\cite{Lomba2017}.

\section{Results}
\label{results}
We will  first  consider
two instances of photoreceptor patterns for low  density
($\rho\sigma^2=0.2$), and moderate density ($\rho\sigma^2=0.8$),
with an interspecies hard core exclusion defined by $\Delta = -0.2$
(i.e. $\sigma_{ij}=0.8\sigma_{ii}$). The coupling factor of the long
range interaction is set to $\Gamma = 5$, and the long-range
cross interactions are set to zero (i.e. $\lambda=0$ in
(\ref{uijlog})).  This means that unlike particles will only interact
via  a pure hard core exclusion. For
comparison we will also show results for
$\lambda=1$ which will only display global hyperuniformity. Partial
densities for each photoreceptor type are $\rho\sigma^2/3$.  

Two
characteristic Monte Carlo snapshots of the low and high density
multihyperuniform systems are presented in Figure \ref{snap3c}.
One can appreciate in the snapshot of \ref{sn3c02}  for
$\rho\sigma^2=0.2$ that photoreceptors of different type
 tend to aggregate in clusters with hetero-coordination. These clusters
 form a low density fluid-like structure, with average inter-cluster
 distances $\approx 3-4\sigma$. When comparing this illustration with
 real representations of bird cone distributions (see Figure 1 in
 Ref.~\cite{Kram2010}) the similarity is evident. At higher densities ($\rho\sigma^2=0.8$)
 packing effects become dominant and clustering is not so apparent,
 but hetero-coordination is still clearly seen in the snapshot of Figure
 \ref{sn3c08}.  The  cluster size distribution (not shown) is monotonously decreasing,
 with no dominant cluster size. This is a consequence of the lack of a
 competing short range attraction that would counteract the long-range
 repulsion  and would thus stabilize finite size clusters, as it was
 the case for $\Delta >0$ in Ref.~\cite{Lomba2018}.

\subsection{Structure factor analysis}
\label{struct}
\begin{figure}[t]
  \includegraphics[width=8cm,clip]{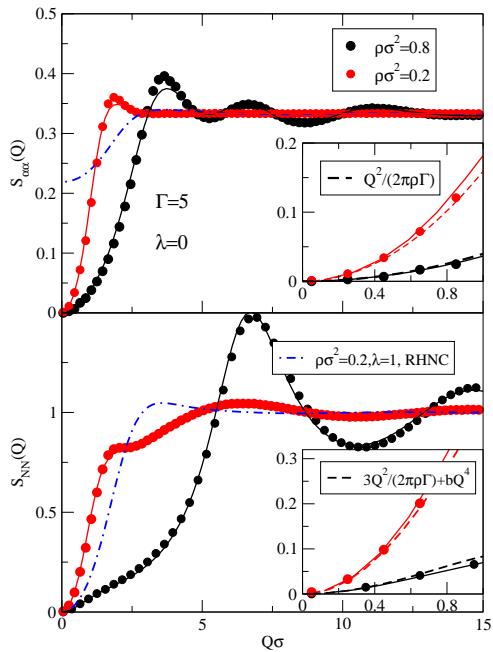}
  \caption{Total and partial  structure  factors of our model
    photoreceptor system displaying multi-hyperuniformity for low and
    moderate number densities (see legend). Solid and dash-dotted curves correspond
    to theoretical (RHNC) calculations, symbols denote Monte Carlo data. Dashed
    curves in the insets represent the low-$Q$ regime derived from
    Eq.(\ref{sijq0}). Partial structure factors correspond to
    correlations between like particles. For comparison we show on blue dash-dotted curves
    the structure factors for a system displaying only global
    hyperuniformity ($\lambda=1$ in Eq.~(\ref{uijlog}).\label{sq3}
  } 
\end{figure}

In
 Figure \ref{sq3} we plot the partial and total structure
 factors corresponding to the systems described above.  The
 multi-hyperuniform character of the system is clearly illustrated by
 their vanishing behavior for 
 low-$Q$. In the insets one can observe that they closely follow the
 asymptotic behavior described by Eqs.~(\ref{sijq0}) and
 (\ref{snnq0}). Theory and simulation agree to a very large extent. 

For comparison we also plot the theoretical results for $\rho\sigma^2=0.2$,
  and $\lambda=1$.  Now, this choice of the long-range cross
  interactions leads to a   globally
 hyperuniform configuration, as confirmed by the behavior of
 $S_{NN}(Q)$ as $Q\rightarrow 0$. In contrast, the partial structure factor does not
 vanish for  $Q\rightarrow 0$ (which rules out multihyperuniformity).
 Given the low density, the result is close to 
 that of an ideal gas, for which $S_{ii}(Q) \approx x_i$ $\forall\;
 Q$. Reducing $\lambda$, which actually implies decreasing (or in our
 present   case, eliminate) the unlike long-range
 repulsive interactions,  induces a certain degree of clustering
 between unlike particles.
 This effect is  visible when comparing  the total structure factor at
 low density (lower graph, red curve in Figure \ref{sq3}) for
 $\lambda=0$ and $\lambda=1$. 
 Only in the case of $\lambda=0$, $S_{\alpha\alpha}(Q)$ exhibits a prepeak  at $Q\sigma \approx
 1.9$. This reflects the presence of clustering with a correlation
 length of $\approx 3.2\sigma$ which we have already qualitatively
 detected in the snapshot of Figure \ref{sn3c02}. In summary, the
 combination of very long-ranged repulsions between like particles
 with non-additive unlike interactions both in the short and long
 range reproduces the features sought for in our minimal statistical
 mechanical model of retina. 

  \begin{figure}[t]
  \includegraphics[width=8cm,clip]{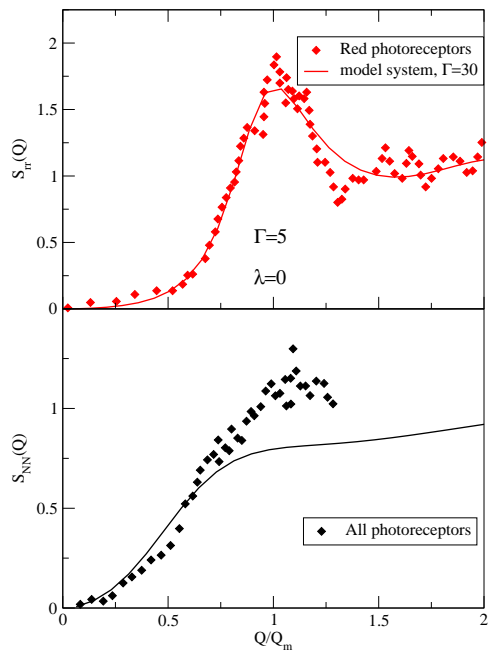}
  \caption{Total and partial structure factor of the symmetric three component plasma with
    negative non-additivity for $\lambda=0$, $\rho\sigma^2=0.8$,
    $\Gamma=30$ compared with those from avian photoreceptors from
    Ref.~\cite{jiao2014}. In the upper curve only the red
    photoreceptors are presented. Solid curves correspond
    to RHNC calculations, symbols to experimental
    data\cite{jiao2014}. The partial structure factor is normalized
    to one and the $Q$ axis is scaled with the position of the
    maximum which is equivalent to adjust $\sigma$ to the effective
    experimental value in the photoreceptor correlations.\label{sqav}
  } 
\end{figure}

\subsection{Mimicking avian retina}

 How do our model results compare with a real structure factor obtained from
 a distribution of avian photoreceptors ?  One must first bear in mind
 that in  bird retina 
  five different types of photoreceptors~\cite{jiao2014} are present
 in  unequal numbers, so in principle our model departs significantly
 from  
 the real situation. Nonetheless, a simple inspection of the
 experimental structure factors presented in
 Figure 9 of Ref.~\cite{jiao2014} indicates that basically all
 photoreceptor species qualitatively  display similar partial structure
 factors.  The global structure factor
  is qualitatively different, with very little structure at low Q
 values. Therefore, it seems reasonable to compare our model system
 with actual experimental results from a qualitative standpoint. To that aim, we have adjusted
 the coupling constant $\Gamma$  of our effective potential to match the results of 
 \cite{jiao2014}. Density is basically coupled to $\Gamma$ (except for
 subtle hard core effects not visible in the low $Q$ behavior of the
 structure factor), so we have set $\rho\sigma^2=0.8$ and kept it
 fixed. As in Ref.~\cite{jiao2014} $Q$ is scaled with the position of
 the structure factor maximum, which sets the length scale to the
 appropriate value in order to ease the comparison. This is equivalent to
 rescaling the data so as to account for  
 the appropriate sizes of the photoreceptors.  We
 observe that the behavior of our simple model depicted in Figure
 \ref{sqav}  agrees qualitatively with the experimental data. As a
 matter of fact, even if in Ref.~\cite{jiao2014} the experimental
 low-$Q$ behavior seems to follow a linear decay instead of the
 $Q^2$ dependence of our model, the quadratic dependence appears acceptable.
  Interestingly, the multiscale packing model also
 proposed  by Jiao and coworkers\cite{jiao2014} displays the same
 quadratic decay. The other salient feature that is observed in Figure
 \ref{sqav} is the lack of  structure of $S_{NN}(Q)$ for low $Q$
 values. This feature is visible both in our simple model (although
 somewhat enhanced) and in the
 experimental data. It is apparent that our minimal model is capable
 of reproducing key features of the spatial patterns display by
 photoreceptors in actual bird retina. 
   \begin{figure*}[t]
   \center
   \subfigure[Photoreceptor model, red cones with $\rho_{red}\sigma^2=0.2/3$]{\includegraphics[width=7.5cm,clip]{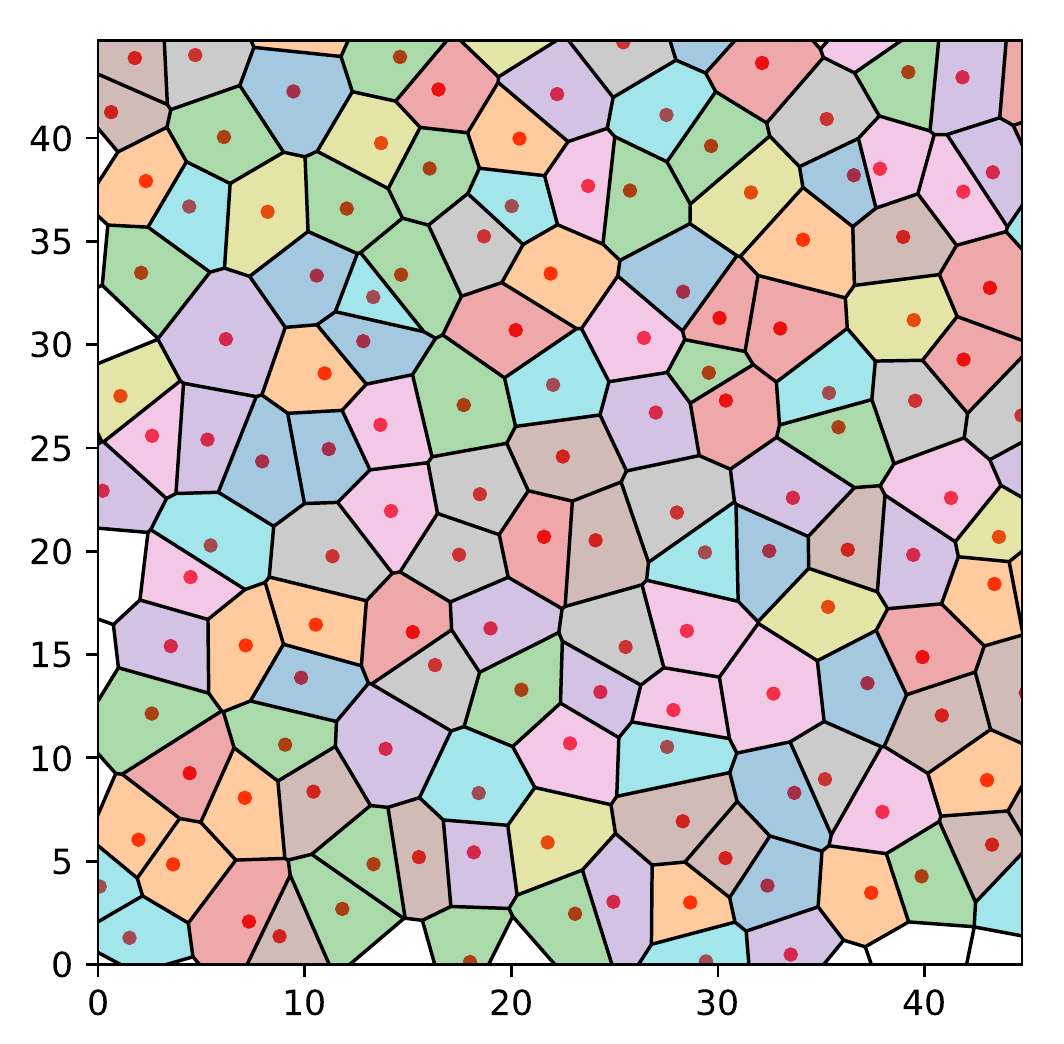}\label{PL02}}\hspace{1cm}
   \subfigure[Photoreceptor model, rgb cones, $\rho\sigma^2=0.2$]{\includegraphics[width=7.5cm,clip]{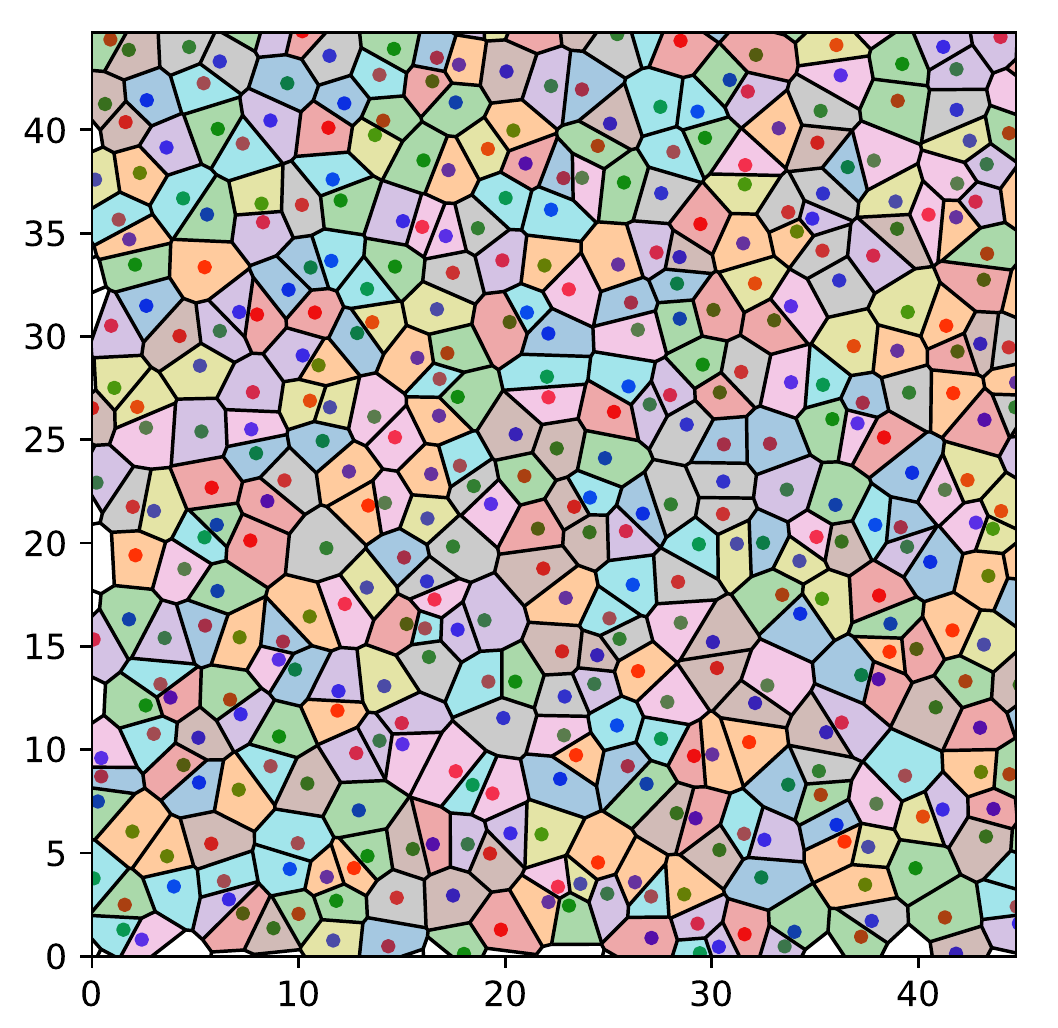}\label{vorfull}}

   \caption{Voronoi
     tessellations corresponding to the red cones (left) of the
     photoreceptor model and all the photoreceptors (right) for total
     density $\rho\sigma^2=0.2$.\label{vorall}}
 \end{figure*}

\subsection{Voronoi analysis}

\begin{figure}[t]
  \includegraphics[width=8cm,clip]{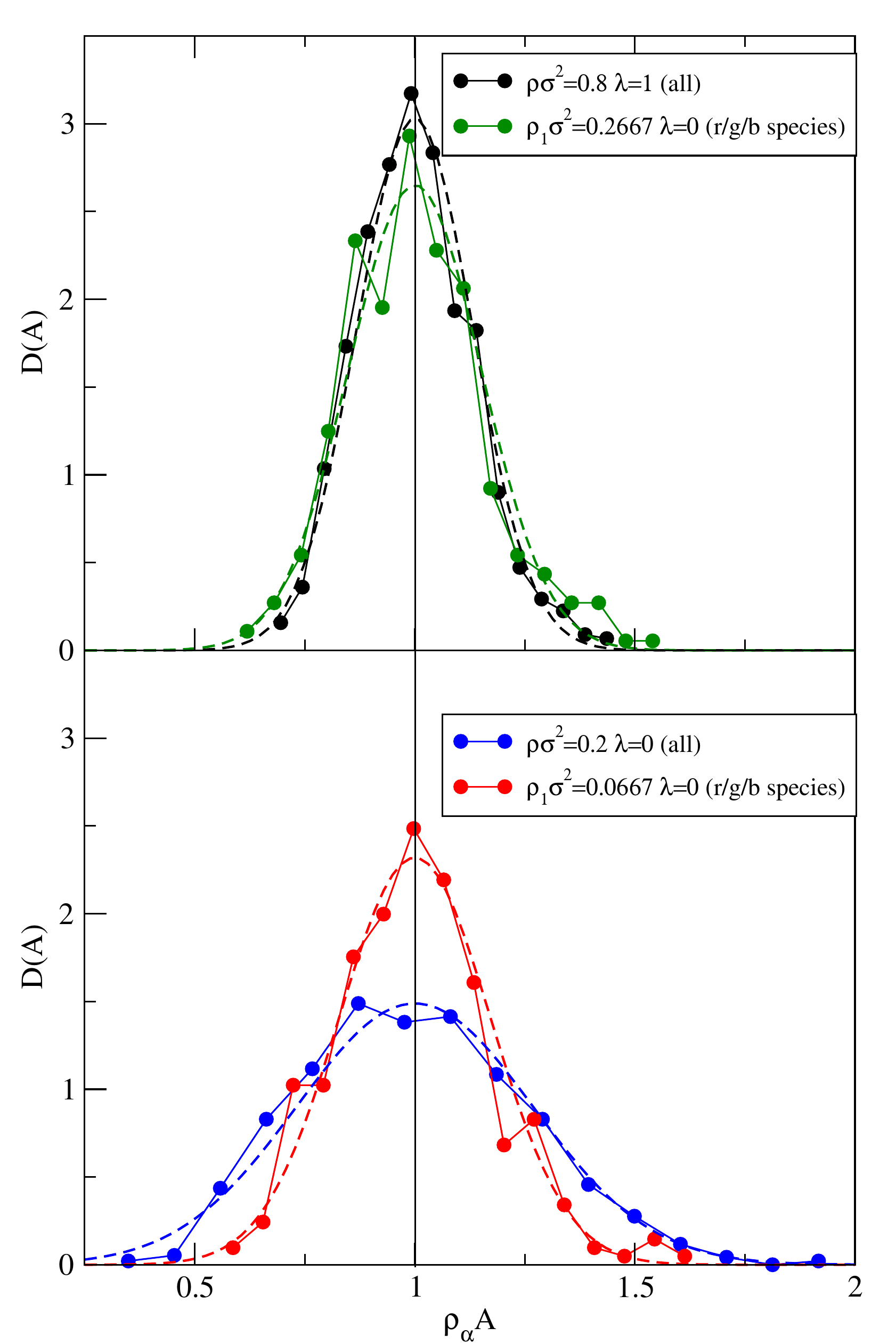}
  \caption{Scaled area distribution of the Voronoi polygons for
    hyperuniform states: high density (upper graph) and low density
    (lower graph) single species and global
    configurations of the three-component plasma. Hyperuniformity
    symmetrizes the area distributions around the value of the square
    regular lattice ($\delta(\rho A -1)$) and the curves follow an
    apparent Gaussian distribution (as shown by the fits represented
    by dashed curves). \label{densg}
  } 
\end{figure}

When thinking of photoreceptors, one must also take into account that
their ability to reproduce an image
is directly related with the area they sample. This
 suggests that a Voronoi analysis of our point
configurations will provide information as to the sampling area
corresponding to  each particle. We have  therefore performed a
characterization of the spatial configurations of our model system
 using  Voronoi tessellations.  We have studied  the corresponding area distribution of the
Voronoi polygons. In order to put these
results in perspective,  we have also performed a corresponding analysis for
purely random two-dimensional point configurations, as well as configurations
obtained from  Molecular Dynamics simulations for  2D fluids of
Lennard-Jones (LJ) particles and LJ particles with added Coulomb
repulsion. In the last instance, the competition between short-range
attractive  and long-range repulsive forces leads to the formation of
stable clusters that nonetheless display hyperuniformity. For these
cases, we have used similar
density conditions and supercritical temperatures ($k_BT/\epsilon =
2.0$, where $k_B$ is Boltzmann's constant and $T$ the absolute
temperature). When referring to LJ results, $\epsilon$ and 
$\sigma$ correspond to the well depth and particle size
respectively. These results are presented in detail in the
Supplementary Information. Upon examination of  Figure  \ref{vorall}, one
can clearly appreciate that  in our model system 
the Voronoi tessellation exhibits a  fairly  regular
distribution of the polygon areas. A similar observation was made by
Legras et al. \cite{Legras2018} when analyzing the cone distribution
in human retina. However, when comparing with tessellations for  LJ
fluids or random distributions (see Figure 1 of the SI) at similar
density, one finds that 
these have a much larger 
dispersion in their areas. This can be more
quantitatively analyzed by examining the normalized area
distributions. These are plotted in Figure
\ref{densg} vs the area scaled with the corresponding particle
densities, $\rho A$. The corresponding figure for random and LJ can be
found in the SI.

 Figure \ref{densg} reveals that our model leads to  the area
distributions that  are symmetrized with
respect to the regular lattice result, $\rho A =1$.  It is interesting
to note that in all cases the curves 
apparently follow a  
Gaussian distribution. In contrast, area distributions for random
configurations and LJ particles at low density follow highly
asymmetric $\Gamma$-distributions, and denser packings of LJ particles
can be fit to log-normal distributions (see SI). It is important to
note that the symmetrization of the area distributions is not a
consequence of hyperuniformity. In Ref. \cite{Zhang2015}, it was found
that certain stealthy hyperuniform patterns led to asymmetric
distributions similar to those of random configurations. This is also
illustrated by the analysis of LJ particles with added Coulombic
repulsions, which form hyperuniform patterns with a strong degree of
clustering. The area distribution of the Voronoi polygons is a
short/medium range property, unlike hyperuniformity which is a
large-scale property.

\begin{figure}[t]
  \includegraphics[width=8cm,clip]{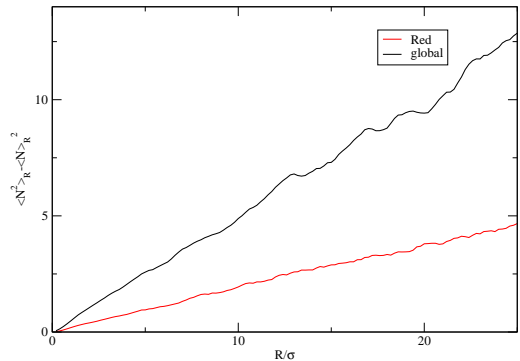}
  \caption{Number variance
$\sigma^2_N(R)$ dependence of sample window radius, $R$, for both the
    global and the single species 
photoreceptor model configurations.\label{nrrgb}
  }
\end{figure}

For our retina model, the marked symmetry of the Voronoi area
distribution is due
to the fact that the interactions are monotonic and repulsive, and
thus tend to produce very regular local environments. This is
illustrated by the uniform linear behavior of number variance
$\sigma^2_N(R)$ for both the global and the single species
photoreceptor configurations in our model, as can be seen in Figure
\ref{nrrgb}. The same linear dependence is found in the experimental
photoreceptor distributions (see Figure 4 in
Ref.~\cite{jiao2014}). Conversely, hyperuniform configurations that
display clustering (and asymmetric 
area distributions of their Voronoi tessellations) present a number
variance with a clear non-monotonic behavior for small sampling
windows. Such is the case of particles with LJ+Coulombic 
interactions as illustrated in Fig. 3 of the SI.

One can then interpret interpret the symmetrization of the area
distribution   in a disordered media as the consequence of the
minimization of repulsive interactions, maximizing the area around each
point in the configuration, and reflecting random deviations from the
the crystalline (ordered hyperuniform) state. This, together
with the strong suppression of long-wavelength density and concentration
fluctuations leads to what could possibly be optimal photoreceptor patterns.

\section{Concluding remarks}
In summary, we have shown that two key features of the experimental
patterns of photoreceptors in bird retina, multi-hyperuniformity and hetero-coordination, can be captured by a simple
model with logarithmic repulsions between like particles and hard core
exclusions with negative non-additivity between the unlike ones. 
The fact that disordered hyperuniform systems represent
topological states of matter sharing fluid and crystal-like properties
makes them the solution of choice when
regular arrangements such as those of arthropod eyes are hampered by
the variability of the  photoreceptors
(e.g. unequal sizes and numbers). Present day bio-inspired optical devices rely on
regular arrangements \cite{Song2013,Garcia2018}. In certain instances,
the combination of
different types of receptors and in different number might be
required compromising the feasibility of regular arrays of receptors.
Disordered multi-hyperuniformity might then offer an alternative to
overcome these difficulties.

A natural extension of this work should be the extension to
non-Euclidean geometries. Steps in this direction can be found in the works of
Meyra et al. \cite{Meyra2019} and Bo\v{z}i\v{c} and
  \v{C}opar \cite{Bozic2019} for spherical surfaces. Actually
   designs on curved surfaces have been already proposed for regular
   arrays in  Ref.~\cite{Song2013}. Disorder hyperuniform systems on
   curved surfaces  might well have a potentially larger impact on technological
  applications. An  analysis along these lines of photoreceptor patterns in
  humans\cite{Legras2018}  might also be of interest to further our
  understanding of our complex visual system. In fact, in
  Ref.~\cite{Legras2018} it was found that human cones tend to
  preserve locally  hexagonal arrangements. A preliminary  Voronoi
  analysis of the area distributions taken from
  Ref.~\cite{Legras2018} show that these are also  relatively
  symmetric. On the other hand, photoreceptor patterns in human retina
  are also dependent on the eccentricity of the
  sampling area, and this considerably complicates the analysis.

 \acknowledgments
 EL  acknowledges the support from the Agencia Estatal de
  Investigación and Fondo Europeo de Desarrollo Regional (FEDER) under
  grant No. FIS2017-89361-C3-2-P. EL and LG  were also supported by the European
  Union’s Horizon 2020 Research and Innovation Staff Exchange
  programme under the Marie 
  Skłodowska-Curie grant agreement No 734276. S.T. was supported by
  the National Science Foundation under Award No. DMR-1714722.  The
  authors are grateful 
  to Prof. Y. Yiao for kindly providing the structure factor data from
  Ref.~\cite{jiao2014}. 
 
%

 \end{document}